\documentclass[proceedings]{JHEP} 

\usepackage{epsfig}                     

\newbox\mybox
\newcommand\fverb{\setbox\mybox=\hbox\bgroup\verb}
\newcommand\fverbdo{\egroup\medskip\noindent\fbox{\unhbox\mybox}\ }
\newcommand\fverbit{\egroup\item[\fbox{\unhbox\mybox}]}

\font\beeg=cmr17 scaled 1600            
\newcommand\init[1]{\setbox\mybox=\hbox{{\beeg #1}~}%
                   \noindent\global\hangindent=\wd\mybox\global\hangafter-2%
                   \sc\smash{\llap {\lower 13.2pt \box\mybox}}}

\def\GSW_sign{}

\newcommand{\ba}{\begin{array}}
\newcommand{\ea}{\end{array}}
\newcommand{\beq}{\begin{equation}}
\newcommand{\eeq}{\end{equation}}
\newcommand{\be}{\begin{equation}}
\newcommand{\ee}{\end{equation}}
\newcommand{\bea}{\begin{eqnarray}}
\newcommand{\eea}{\end{eqnarray}}

\def\bra{\langle}
\def\ket{\rangle}

\def\a{\alpha}
\def\b{\beta}
\def\g{\gamma}

\def\l{\lambda}
\def\m{\mu}
\def\n{\nu}
\def\G{\Gamma}

\def\to{\rightarrow}

\title{Theory of Rare B Meson Decays}

\author{C. Greub\thanks{Work partially supported by Schweizerischer 
        Nationalfonds. {\bf BUTP-99/22}}\\
        Institut f\"ur theoretische Physik, Universit\"at Bern, CH-3012 Bern\\
        E-mail: \email{greub@itp.unibe.ch}}

\conference{Heavy Flavours 8, Southampton, UK, 1999} 

\abstract{I review the NLO QCD calculations
of the branching ratio for
$B \to X_s \gamma$ in the SM. Including  the leading electromagnetic
corrections, one obtaines $BR(B \to X_s \gamma)=(3.32 \pm 0.30) 
\times 10^{-4}$. Confronting theory with the newest data, an updated value
for $|V_{ts}|$ is obtained: $|V_{ts}|=0.037 \pm 0.007$.
Theoretical progress on the photon energy spectrum is also discussed. The
inclusive FCNC semileptonic decays in 
the SM are briefly summarized. Furthermore, 
$B \to X_s \gamma$ is considered in 2HDMs and in different SUSY scenarios.
QCD corrections are shown to be crucial.}

\begin{document} 

\maketitle 

\section{Introduction}
In the Standard model (SM), rare $B$ meson decays
are induced 
by one-loop diagrams, where $W$ bosons and up-type quarks are exchanged.  
In many extensions of the SM,
there are additional contributions, 
where the SM particles in the loop  are  
replaced by nonstandard ones,
like charged Higgs bosons, gluinos, charginos etc. 
Being induced also at the one loop-level, the new physics
contributions are not necessarily suppressed relative to the SM one.
The resulting sensitivity for nonstandard effects implies the possibility
for an indirect observation of new physics, or allows to put limits
on the masses and coupling parameters of the new particles.
A general overview, where many $B$ physics observables are investigated w.r.t.
new physics, was given at this conference by A. Masiero \cite{Masiero}.
Concerning the new physics aspects of rare $B$ decays, I concentrate
in this article (see sections 3 and 4) on recent 
calculations of the branching ratio for the process $B \to X_s \gamma$
in a general class of two-Higgs-doublet models (2HMDs) 
\cite{BG98,Giudice97}
and in  
supersymmetric scenarios \cite{Giudice98,Misiak99,BGHW99},
stessing in particular the importance of
leading- (LO) and next-to-leading (NLO) QCD corrections.

However, also in the absence of new physics, rare $B$ decays
are very important; they can be used for the determination   
of various CKM matrix elements, occurring in the SM Lagrangian. 
To
extract these parameters from data, it is crucial that the corresponding
decay rates are reliably calculable. An important class of such decays
are the inclusive rare $B$ decays, like $B \to X_s \gamma$, 
$B \to X_s \ell^+ \ell^-$, 
$B \to X_s \nu \bar{\nu}$, 
$B \to X_d \gamma$,
$B \to X_d \ell^+ \ell^-$, 
$B \to X_d \nu \bar{\nu}$, 
which are sensitive to the CKM
matrix elements $|V_{ts}|$ and $|V_{td}|$, respectively. 
In contrast to the corresponding exclusive channels, these 
inclusive decay modes are theoretically cleaner, in the sense 
that no specific model is needed
to describe the final hadronic state. 
Nonperturbative effects in the inclusive modes are well 
under control due to heavy quark effective theory. 
For example, the decay width $\G(B \to X_s \gamma)$ 
is well approximated by the partonic decay rate
$\G(b\to X_s \gamma)$ which can be
analyzed in renormalization group improved perturbation theory. 
The class of 
non-perturbative effects which scale like $1/m_b^2$
is expected to be well below $10\%$ \cite{Falk}. 
This numerical
statement  also holds 
for the non-perturbative contributions
which scale like $1/m_c^2$ \cite{Voloshin,BIR}.

The framework and the NLO theoretical results 
for the branching ratio of the decay $B \to X_s \gamma$ are discussed 
in section 2.1; the photon energy spectrum and the
partially integrated branching ratio for this process are reviewed in
section 2.2;
an updated value for the CKM matrix element
$|V_{ts}|$, extracted  from the most recent measurements of
$BR(B \to X_s \gamma)$ and the corresponding calculations, where also
the leading electromagnetic corrections are included, is 
given in section 2.3; the other inclusive rare decays mentioned above,
are briefly discussed in section 2.4.

The exclusive analogues,  $B \to K^* \gamma$, 
$B \to K^{(*)} \ell^+ \ell^-$,  
$B \to \rho \gamma$,
$B \to \rho \ell^+ \ell^-$ etc., require the
calculation of form factors. As the QCD sum rule calculations 
and the lattice results 
for these form factors 
were summarized by V. Braun \cite{Braun},
I do not discuss these decays in the following.

Finally, there is the class of non-leptonic two-body decays, like
$B \to \pi \pi$, $B \to K \pi$;
the theoretical status of these processes was discussed by L. Silvestrini
\cite{Silvestrini},
and new  CLEO results were presented by D. Jaffe \cite{Jaffe}.

\section{Inclusive rare $B$ meson decays in the SM}
\subsection{$BR(B \to X_s \gamma)$ at NLO precision}
Short distance QCD effects 
enhance the partonic decay rate $ \Gamma(b \to s \gamma)$ 
by more than a factor of two.
Analytically, these QCD corrections contain large logarithms of the form 
$\alpha_s^n(m_b) \, \log^m(m_b/M)$,
where $M=m_t$ or $M=m_W$ and $m \le n$ (with $n=0,1,2,...$).
In order to get a reasonable prediction for 
the decay rate, it is evident that one has  to resum at least
the leading-log (LO) series ($m=n$).  
As the error of the LO result \cite{counterterm}
was  dominated by a large renormalization scale dependence 
at the $\pm 25\%$ level, it became clear that even the NLO terms
of the form $\a_s(m_b) \, \left(\a_s^n(m_b) \, \ln^n (m_b/M)\right)$
have to be taken into account systematically.

To achieve the necessary resummations, one usually contructs in a first
step
an effective low-energy theory and then resums the large logarithms
by renomalization group techniques. The low energy theory is
obtained by integrating out the
heavy particles which in the SM are the top quark and the $W$-boson. 
The resulting effective Hamiltonian relevant for $b \to s \gamma$ 
in the SM and many of its extensions reads 
\begin{equation}
\label{heffW}
H_{eff}^W(b \to s \gamma)
       = - \frac{4 G_{F}}{\sqrt{2}} \, \lambda_{t} \, \sum_{i=1}^{8}
C_{i}(\mu) \, O_i(\mu) \quad , 
\end{equation}
where $O_i(\m)$ are local operators consisting of light fields,
$C_{i}(\mu)$ are the corresponding Wilson coefficients,
which contain the complete top- and $W$- mass dependence,
and $\lambda_t=V_{tb}V_{ts}^*$ with $V_{ij}$ being the
CKM matrix elements. The CKM dependence globally factorizes,
because we work in the approximation $\l_u=0$.

Retaining only operators up to dimension 6
and using the equations of motion, one arrives at the following basis 
\bea
\label{operators}
O_1 &=& \left( \bar{c}_{L \b} \g^\m b_{L \a} \right) \,
        \left( \bar{s}_{L \a} \g_\m c_{L \b} \right)\,, \nonumber \\
O_2 &=& \left( \bar{c}_{L \a} \g^\m b_{L \a} \right) \,
        \left( \bar{s}_{L \b} \g_\m c_{L \b} \right) \,,\nonumber \\
O_7 &=& \frac{e}{16\pi^{2}} \, \bar{s}_{\a} \, \sigma^{\m \n}
      \, (m_{b}(\mu)  R) \, b_{\a} \ F_{\m \n} \,,
        \nonumber \\
O_8 &=& \frac{g_s}{16\pi^{2}} \, \bar{s}_{\a} \, \sigma^{\m \n}
      \, (m_{b}(\mu)  R) \, \frac{\l^A_{\a \b}}{2} \,b_{\b}
      \ G^A_{\m \n} \quad .
\eea
As the Wilson coefficients of the QCD penguin
operators $O_3,...,O_6$ are small, we do not list them here.

It is by now well known that a consistent calculation for 
$b \to s \gamma$ at LO (or NLO) precision requires three steps:
\begin{itemize}
\item[{\it 1)}] 
a matching calculation of the full standard model theory 
with the effective theory at the scale $\mu=\mu_W$ 
to order $\alpha_s^0$ (or $\alpha_s^1$) for the Wilson coefficients, 
where  $\mu_W$ denotes a scale of order $M_W$ or $m_t$;
\item[{\it 2)}]  
a renormalization group evolution of the Wilson coefficients
from the matching scale $\mu_W$ down to the low scale
$\mu_b=O(m_b)$,
using the anomalous-dimension matrix to order $\alpha_s^1$ 
(or $\alpha_s^2$);
\item[{\it 3)}]   
a calculation of the matrix elements of the operators at the scale 
$\mu = \mu_b$  to order $\alpha_s^0$ (or $\alpha_s^1$). 
\end{itemize}

At NLO precision, all three steps are rather involved:
The most difficult part in Step 1 is the 
order $\alpha_s$ matching of the dipole operators $O_7$ and $O_8$. 
Two-loop diagrams, both in the full- and in the effective theory
have to be worked out. This matching calculation was first performed
by Adel and Yao \cite{Adel} and then  
confirmed by Greub and Hurth \cite{GH}, using a different method. Later,
two further recalculations of this result were presented
\cite{Giudice97,INFRARED}.
The order $\a_s^2$ anomalous matrix (Step 2) has been 
worked out
by Chetyrkin, Misiak and M\"unz \cite{Mikolaj}. The extraction of
certain elements in this
matrix involves the calculation of pole parts of three loop diagrams. 
Step 3  consists of Bremsstrahlung contributions and virtual
corrections. The Brems- strahlung corrections
were worked out some time ago by Ali and Greub \cite{AG91} and have
been confirmed and extended by Pott \cite{Pott}. An  
analysis of the virtual two loop corrections was presented
by Greub, Hurth and Wyler \cite{GHW}. 

Combining the NLO calculations of these 3  steps, 
leads to the following NLO QCD prediction  
for the branching ratio \cite{BG98}: 
\begin{equation}
 BR(B \to X_s \gamma) = \left(
       3.57  \pm^{\,0.01}_{\,0.12} 
             \pm^{\,0.00}_{\,0.08} 
             \pm^{\,0.29}_{\,0.27} 
         \right) \times 10^{-4}  \,. 
\label{NLOQCD}
\end{equation}
The central value is obtained for 
$\mu_b=4.8\,$GeV, $\mu_W = m_W$ and the central values of the input 
parameters listed in  \cite{BG98}.
The first error is obtained by varying $\mu_b$ in the interval
$[2.4,9.6]\,$GeV, the second one by varying the matching scale $\mu_W$
between $m_W$ and $m_t$; the third error reflects the uncertainties
in the various input parameters. Similar results were also obtained
in refs. \cite{Giudice97,INFRARED}.

We should mention that in the result (\ref{NLOQCD}) also power corrections
are included: there are $1/m_b^2$ corrections \cite{Falk}, 
whose impact on $BR(B \to X_s \gamma)$ is at the $1\%$ level, as well as
nonperturbative contributions from $c\bar c$ intermediate states 
which scale with $1/m_c^2$.  
Detailed investigations \cite{Voloshin,BIR}
show that these $1/m_c^2$ corrections enlarge
the branching ratio by $\sim 3\%$.

After the QCD analyses, several papers appeared where different 
classes of  electroweak corrections~\cite{CM,KN,STRUMIA,Baranowski} to 
$BR(B \to X_s\gamma)$ were considered. 
In~\cite{STRUMIA}, corrections to the Wilson coefficients 
at the matching scale due to the top quark and the neutral 
Higgs boson were calculated and found to be negligible. 
The analysis ~\cite{CM} concluded that the most
appropriate value of $\alpha_{em}^{-1}$ to be used for this 
problem is the fine structure constant 
$\alpha^{-1} = 137.036$ instead of the value 
$\alpha_{em}^{-1} = 130.3\pm 2.3$ 
previously used. In~\cite{KN,Baranowski}, the leading 
logarithmic QED corrections of the form 
$\alpha \log ({\mu_W}/{\mu_b}) \left(
\alpha_s \log ({\mu_W}/{\mu_b}) \right)^n$ (with resummation 
in $n$) were given. 

In ref. \cite{BG98add} we updated the result in eq. (\ref{NLOQCD}), 
by including the class of 
QED corrections presented in~\cite{KN}; we then obtained  
\begin{equation}
BR(B \to X_s\gamma)
 = \left(
  3.32  \pm^{\,0.00}_{\,0.11} 
        \pm^{\,0.00}_{\,0.08} 
        \pm^{\,0.26}_{\,0.25} 
   \right) \times 10^{-4}  \,. 
\label{NLOQCDQED}
\end{equation}
The bulk of the change with respect to the value 
in eq. (\ref{NLOQCD}) is due to the different 
value of $\alpha_{em}^{-1}$
used. 

A remark concerning the error due to the variation of the
low scale $\mu_b$ in the results (\ref{NLOQCD}) and 
(\ref{NLOQCDQED}) is in order here: As it will be discussed in more detail
 in section 3, it was realized in \cite{BG98}
that in multi Higgs doublet models the QCD corrections in certain 
regions of the parameter space are much larger than in the SM. 
As a consequence, the dependence on the scale $\mu_b$ of 
$BR(B \to X_s \gamma)$ is also larger. Later, Kagan and Neubert
pointed out very explicitly in their analysis \cite{KN} that the
scale dependences in individual contributions to the branching ratio
in the SM
are larger than their combined effect, due to accidental cancellations. 
They suggest 
that one should add the scale uncertainties from the individual
contributions in quadrature, in order to get a more reliable estimate of the
 truncation error. Their estimate for the $\mu_b$ dependence 
of $BR(B \to X_s \gamma)$ is
$\pm 6.3\%$, i.e., more than twice the naive estimate. 
The total error in (\ref{NLOQCDQED}), however, is dominated by
parametric uncertainties and therefore gets increased only 
marginally when  
using this more conservative estimate of the scale uncertainties.

The measurement of $BR(B \to X_s \gamma)$ 
by the ALEPH collaboration at LEP \cite{BRALEPH} 
\begin{equation}
\label{BRALEPH}
BR(B \to X_s \gamma) = (3.11 \pm 0.80 \pm 0.72) \times 10^{-4} 
\end{equation}
and by the CLEO collaboration at CESR \cite{CLEOneu}
\begin{equation} 
\label{BRCLEO}
BR(B \to X_s \gamma) = (3.15 \pm 0.35 \pm 0.32 \pm 0.26) \times 10^{-4} 
\end{equation}
are in good agreement with the NLO calculation (\ref{NLOQCDQED}),
where the most important electromagnetic corrections are included.

\subsection{Partially integrated branching ratio in $B \to X_s \gamma$}
The photon energy spectrum of the partonic decay $b \to s \gamma$
is a delta function, concentrated at $\sim (m_b/2)$, 
when the b-quark decays at rest. This delta function gets smeared when
considering the inclusive photon energy spectrum
from a $B$ meson decay. There is a perturbative contribution to this smearing, 
induced by
the Bremsstrahlung process $b \to s \gamma g$ \cite{AG91,Pott},
as well as a non perturbative
one, which is due to the Fermi motion of the decaying $b$ quark in 
the $B$ meson. 

For small photon energies, the $\gamma$-spectrum from $B \to X_s \gamma$ is
completely overshadowed by background processes, 
like $b \to c \bar{u} d \gamma$ and $b \to u \bar{u} d \gamma$.
This background falls off very rapidly with increasing photon
energy, and becomes small for $E_\gamma > \sim 2$ GeV \cite{AG92}.
 This implies
that only the partial branching ratio
\begin{equation}
BR(B \to X_s \gamma)(E_\gamma^{min}) = \int_{E_\gamma^{min}}^{E_\gamma^{max}}
 \frac{dBR}{dE_\gamma}
dE_\gamma
\end{equation}
can be directly measured, with $E_\gamma^{min}=O(2)$ GeV.
Recently, CLEO was able to reduce $E_\gamma^{min}$ from 2.2 GeV to 2.1 GeV
\cite{CLEOneu}.
To determine from such a measurement the full branching ratio for
$B \to X_s \gamma$,
one has to know from theory the fraction $R$ of the $B \to X_s \gamma$
events with photon energies above $E_\gamma^{min}$.  
Based on calculations by Ali and Greub \cite{AG91} 
of the photon energy spectrum within
the Fermi motion model by Altarelli et al.
\cite{ACCMM}, CLEO used the value
$R= 85\% - 94\%$ \cite{CLEOneu} 
in order to determine $BR(B \to X_s \gamma)$ from
the measured partial branching ratio.

A modern way - based on first principles -  implements 
the Fermi motion in the framework
of the heavy-quark expansion. When probing the spectrum
closer to the endpoint, the OPE breaks down, and the leading twist
non-perturbative corrections must be resummed into the $B$ meson structure
function $f(k_+)$ \cite{shape}, where $k_+$ is the
light-cone momentum of the $b$ quark in the $B$ meson.
The physical spectrum is then obtained by
the convolution 
\begin{equation}
\frac{d\Gamma}{dE_\gamma} = \int_{2E_\gamma-m_b}^{\bar{\Lambda}} dk_+ f(k_+)
\frac{d\Gamma_{part}}{dE_\gamma}(m_b^*) \quad ,
\end{equation}
where $(d\Gamma_{part}/dE_\gamma)(m_b^*)$ 
is the partonic
differential rate, written as a function of the ``effective mass''
$m_b^*=m_b+k_+$. 
The function $f(k_+)$ has support in the range $-\infty < k_+ <
\bar{\Lambda}$, where $\bar{\Lambda}=m_B-m_b$ 
in the infinite mass limit. This implies that
the addition of the structure function 
moves the partonic endpoint
of the spectrum from $m_b/2$ to the physical endpoint $m_B/2$.
While the shape of the function $f(k_+)$ is unknown, the first few
moments $A_n = \int dk_+ \, k_+^n f(k_+)$ are known: $A_0=1$, $A_1=0$
and $A_2 = -\lambda_1/3$. The values of $\bar{\Lambda}$ and 
$\lambda_1$ are not calculable analytically and have to be extraced from
experiments or calculated on the lattice. A recent analysis gives
$(\bar{\Lambda},\lambda_1)=(0.39 \pm 0.11 \mbox{GeV}, -0.19 \mp 0.10 
\mbox{GeV}^2)$ \cite{Gremm}. 
As $A_n$ ($n>2$) are poorly known, several Ans\"atze
were used for $f(k_+)$; e.g. Neubert and Kagan \cite{KN} 
used $f(k_+)= N (1-x)^a
e^{(1+a)x}$, with $x=k_+/\bar{\Lambda}$. Taking into account the constraints 
from $A_0$, $A_1$ and $A_2$, the independent parameters in this Ansatz
can be chosen to be $m_b$ and $\lambda_1$. 
As shown in \cite{KN}, the uncertainty of 
$m_b$ dominates the error of the partial branching ratio. 
\FIGURE{\epsfig{file=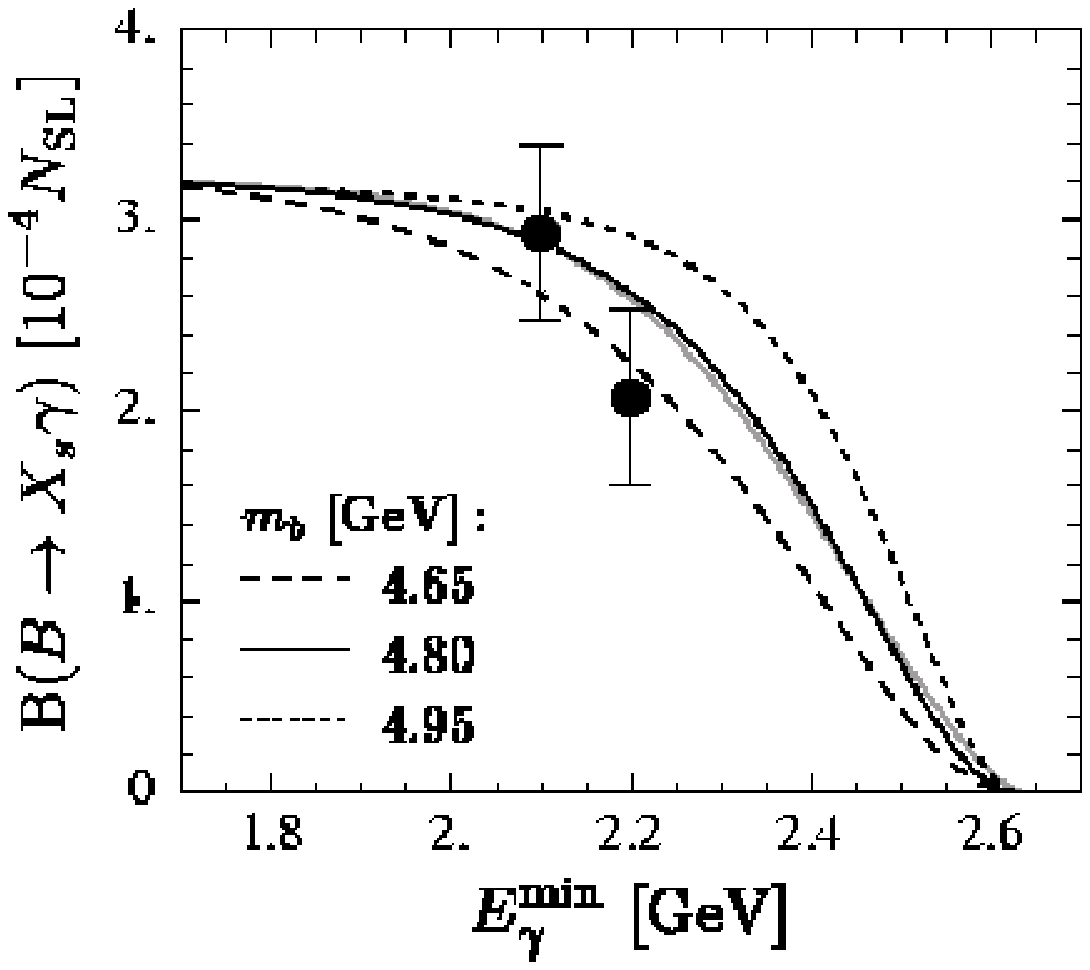,height=5.00cm,width=7.00cm}%
        \caption{Partially integrated branching ratio as a function of
                 the energy cutoff $E_\gamma^{min}$; 
                 Figure taken from Neubert and Kagan \cite{KN}.}
        \label{fig:neubert}}
In Fig.~\ref{fig:neubert}
the partial branching ratio is shown for the relevant range of $m_b$
as a function of $E_\gamma^{min}$, keeping $\lambda_1/\bar{\Lambda}^2$ fixed.
For comparison, the data point 
$BR(B \to X_s \gamma)=(2.04 \pm 0.47)\times 10^{-4}$, 
obtained in the original
CLEO analysis with a cutoff at 2.2 GeV \cite{CLEOold}, 
as well as the new data point
$BR(B \to X_s \gamma)=(2.97 \pm 0.33 \pm 0.30 \pm 0.21) \times 10^{-4}$, 
corresponing to a cutoff at 2.1 GeV \cite{CLEOneu}, 
are also shown in Fig. \ref{fig:neubert}.
We would like to stress that it would be very welcome, if the cutoff
could be pushed down to 2.0 GeV, because the theoretical uncertainity 
drastically becomes smaller, as seen in Fig. \ref{fig:neubert}.

We should add that in the endpoint region also the perturbative
contributions, mentioned at the beginning
of this section, could become problematic in principle, 
due to the presence of threshold logs,
$\log(1-2E_\gamma/m_b)$. These logarithms have recently been resummed 
up to next-to-leading logarithmic precision \cite{Rothstein}.
The authors of this paper conclude that for 
the present energy cut at 2.1 GeV, these
threshold logs do not form a dominant sub-series and therefore their
resummation is not necessary for predicting the decay rate.
Recently, also the BLM 
type corrections of the order $\alpha_s^2 \beta_0$
to the photon energy spectrum were calculated \cite{Ligeti}. 
The result was  used to extract a value for $\bar{\Lambda}$
from the average $\bra 1-2E_\gamma/m_B \ket$, and a value for $\lambda_1$
from  $\bra (1-2E_\gamma/m_B)^2 \ket$. According to this analysis, the CLEO
data in the region $E_\gamma > 2.1$ GeV implies the central value
$\bar{\Lambda}_{\alpha_s} \simeq 390$ MeV and
$\bar{\Lambda}_{\alpha_s^2 \beta_0} \simeq 270$  MeV at
order $\alpha_s$ and $\alpha_s^2 \beta_0$, respectively. This anlysis
was somewhat critisized in ref.~\cite{Rothstein}, by pointing out that
other $\alpha_s^2$ terms could be  larger than the BLM terms.

\subsection{$|V_{ts}|$ form $B \to X_s \gamma$}
Instead of making a
prediction for $BR(B \to X_s \gamma)$,
one can use the NLO calculation to extract
the CKM combination $|V_{tb} V_{ts}^*|/|V_{cb}|$ from the measurements; 
in turn, one can determine 
$|V_{ts}|$, by making use of the relatively well  known
CKM matrix elements $V_{cb}$ and $V_{tb}$.  
Using the CLEO (\ref{BRCLEO}) and
ALEPH data (\ref{BRALEPH}), one obtains \cite{Montpellier98}
\begin{equation}
\label{CKMcombine}
 \frac{|V_{ts}^* V_{tb}|}{|V_{cb}|} = 0.93 \pm 0.09_{exp.}  \pm 0.03_{th} \, .
\end{equation}
Using
$|V_{tb}|=0.99 \pm 0.15$ from the CDF 
measurement and  $|V_{cb}|=0.0393 \pm 0.0028$  
extracted from semileptonic $B$ decays, one obtains
\begin{equation}
\label{vts}
|V_{ts}|=0.037 \pm 0.007,
\end{equation}
where all the errors were added in quadrature. This  is probably
the most direct determination of this CKM matrix element,
as the measurement of $t \to s W^+$ seems to be difficult.
With an improved measurement of $BR(B \to X_s \gamma)$
and $V_{tb}$, one expects to reduce the present error on $|V_{ts}|$
by a factor of 2 or even more.

\subsection{$B \to X_d \gamma$, $B \to (X_s,X_d) \ell^+ \ell^-$ and
$B \to (X_s,X_d) \nu \bar{\nu}$ in the SM}
The decay $B \to X_d \gamma$ can be treated in a similar way as
$B \to X_s \gamma$ \cite{AAG98}.
The only difference is that
$\lambda_u$ for $b \to d \gamma$ is not small relative to
$\lambda_t$ and $\lambda_c$; therefore, also
the current-current operators $O_1^u$ and $O_2^u$, 
weighted by $\lambda_u$, contribute.
Unfortunately, these operators induce long-distance 
contributions to $B \to X_d \gamma$, 
which at present only can be estimated using models. 
In ref.~\cite{AAG98}, these long-distance effects were absorbed into
the theoretical error.

Using $\mu_b=2.5$ GeV and the central values of the input
parameters, the analysis in ref.~\cite{AAG98} yields 
a difference between the LO and NLO predictions
for $BR(B \to X_d \gamma) $ of 
$\sim 10\%$, increasing the branching ratio in the NLO case.
For a fixed value of the CKM-Wolfenstein parameters 
$\rho$ and $\eta$, the theoretical uncertainty of the average branching
ratio $\bra BR(B \to X_d \gamma) \ket$ of the decay $B \to X_d \gamma$
and its charge conjugate $\overline{B} \to \overline{X_d} \gamma$ is:
$\Delta \bra BR(B \rightarrow X_d \gamma)\ket/ 
\bra BR(B \rightarrow X_d \gamma) \ket   
= 
\pm (6-10)\%$. 
Of particular theoretical interest for constraining $\rho$ and $\eta$
is the ratio of the
branching ratios, defined as
\begin{equation}
\label{dsgamma}
R(d\gamma/s\gamma) \equiv \frac{\bra BR(B \to X_d \gamma) \ket}
                           {\bra BR(B \to X_s \gamma) \ket},
\end{equation}
in which a good part of the theoretical uncertainties cancels. 
Varying the CKM-Wolfenstein parameters $\rho$ and $\eta$ in the range
$-0.1 \leq \rho \leq 0.4$ and $0.2 \leq \eta \leq 0.46$ and taking into
account other parametric dependences, the 
results (without electroweak corrections) are
\begin{eqnarray}
\label{summarybrasy}
6.0 \times 10^{-6} &\leq &
BR(B \rightarrow X_d \gamma)   \leq 2.6 \times 10^{-5}~, \nonumber\\
0.017 &\leq & R(d\gamma/s\gamma) \leq 0.074~.\nonumber
\end{eqnarray}
Another observable, which is also sensitive to the CKM parameters
$\rho$ and $\eta$, is the CP rate asymmetry $a_{CP}$, defined as
\begin{equation} 
a_{CP} = 
\frac{\Gamma(B \to X_d \gamma)-\Gamma(\overline{B} \to \overline{X_d} \gamma)}{
      \Gamma(B \to X_d \gamma)+\Gamma(\overline{B} \to \overline{X_d} \gamma)}
    \, .
\end{equation} 
Varying $\rho$ and $\eta$ in the range specified above, we
obtained $7\% \le a_{CP} \le 35\%$ \cite{AAG98}.
We would like to point out that $a_{CP}$ is at the moment only available to
LO precision and therefore suffers from a relatively large renormalization
scale dependence.

A measurement of the semileptonic FCNC decays
$B \to X_s \ell^+ \ell^-$ and $B \to X_d \ell^+ \ell^-$,
below the $J/\psi$- and above the $\rho,\omega$-resonance regions
in the dilepton invariant mass, can also be used to extract $|V_{ts}|$
and $|V_{td}|$, respectively. In this context, these decays and the related
ones, $B \to X_s \nu \bar{\nu}$ and $B \to X_d \nu \bar{\nu}$, were
discussed some time ago \cite{AGM}. 
The decays $B \to (X_s,X_d) \nu \bar{\nu}$
are practically free of long-distance contributions \cite{BIR}
and the renormalization
scale dependence of these decay rates has also been brought under control
\cite{BB}. 
Hence, these decays are theoretically remarkably clean but, unfortunately,
they are difficult to measure.
The ALEPH collaboration has searched for the decay
$B \to X_s \nu \bar{\nu}$, setting an upper bound
$BR(B \to X_s \nu \bar{\nu})<7.7 \times 10^{-4}$ (at 90\% C.L.) 
\cite{ALEPHnu}, which is a factor 20 away from the SM expectations
\cite{BB}.

In contrast, the prediction of the decay rate 
$\Gamma(B \to X_s \ell^+ \ell^-)$ still
suffers from many uncertainties. 
The most important ones are due to intermediate $c \bar{c}$ states.
Because of the non perturbative nature of these states, the differential
spectrum can be only roughly
estimated when the invariant mass $m_{\ell^+ \ell^-}$ is not sufficiently
below $m_{J/\psi}$. However, for low
$\hat{s}=m_{\ell^+ \ell^-}^2/m_b^2$, a relatively
precise determination of the spectrum is possible using perturbative
methods only, up to calculable HQET corrections. The dominant HQET corrections
were evaluated 
\footnote{ 
For an overview of different treatments of
the $1/m_b^2$ and $1/m_c^2$ corrections in $B \to (X_s,X_d) \ell^+ \ell^-$,
and of the $\Lambda_{QCD}^2/q^2$ terms, generated by the $u$-quark loops
in $B \to X_d \ell^+ \ell^-$, we refer to \cite{AH98}.}
and found to be smaller than 6\% for
$0.05 < \hat{s} < 0.25$.
Therefore, the $B \to X_s \ell^- \ell^+$
rate integrated over this region of $\hat{s}$ should be perturbatively
predictable
as precisely as $\Gamma(B \to X_s \gamma)$, i.e. up to about 10\%
uncertainty. However, 
the presently available NLO QCD corrections \cite{MisiakZH,BM} have not yet
reached this precision. The formally LO term is suppressed, which makes it 
as small as some of the NLO contributions. Consequently, some NNLO terms
still can be large. The NNLO program for this decay was recently started
\cite{Misiaknew}, by calculating the two-loop matching conditions for
all the operators relevant for $B \to X_s \ell^+ \ell^-$. The improved
matching allows to remove an important ($\sim \pm 16\%$) matching
scale uncertainty for $BR(B \to X_s \ell^+ \ell^-)$ 
in the mentioned region of $\hat{s}$, leading to 
$ BR(B \to X_s \ell^+\ell^-) = 1.46 \times 10^{-6}$. A remaining
perturbative uncertainty of about 13\%, due to the unknown two-loop matrix
elements of the four-quark operators, was also estimated in 
ref. \cite{Misiaknew}.

\section{$B \to X_s \gamma$ in generalized two-Higgs Doublet models}
Two Higgs Doublet Models (2HDMs) are conceptually among the simplest
extensions of the SM. 
Studies of
$BR(B \to X_s \gamma)$  in these models can already test
whether  the observed high accuracy of the NLO SM result  
is a generic feature of NLO calculations \cite{BG98,BG98add} 
or a rather particluar
one, valid for the SM only.
Such studies can obviously provide also important indirect bounds on the
new parameters contained in these models. 

The well--known Type~I and Type~II models are particular examples of
2HDMs, in which the same or the two different Higgs fields couple to
up-- and down--type quarks.  The second one is especially important
since it has the same couplings of the charged Higgs $H^+$ to fermions
that are present in the Minimal Supersymmetric Standard Model
(MSSM). The couplings of the neutral Higgs to fermions have
important differences from those of the MSSM~\cite{KRA,BD}. However, since
beside the $W$, only charged Higgs bosons mediate the decay $B \to X_s \gamma$
when additional Higgs doublets are present, the predictions of 
$BR(B \to X_s \gamma)$
in a 2HDM of Type II give, at times, a good approximation of the value
of this branching ratio in some supersymmetric models~\cite{MMM}.

It is implicit in our previous statements that we do not consider
scenarios with tree--level flavour changing couplings to neutral Higgs
bosons.  We do, however, generalize our class of models to accommodate
Multi--Higgs Doublet models, provided only one charged Higgs boson
remains light enough to be relevant for the process $B \to X_s \gamma$.  This
generalization allows a simultaneous study of different models,
including Type~I and Type~II, by a continuous variation of the
(generally complex) charged Higgs couplings to fermions.  It allows
also a more complete investigation of the question whether the
measurement of $BR(B \to X_s \gamma)$ 
closes the possibility of a relatively light
$H^\pm$ not embedded in a supersymmetric model.

We will show that the NLO QCD corrections to the Higgs contributions
to $BR(B \to X_s \gamma)$ are much larger than the corresponding corrections
to the SM contribution \cite{BG98}, irrespectively of the value of the
charged Higgs couplings to fermions.  
This feature remains undetected in Type~II models, where the
SM contribution to $BR(B \to X_s \gamma)$ 
is always larger than, and in phase with,
the Higgs contributions. In this case, a
comparison between theoretical and experimental results for 
$BR(B \to X_s \gamma)$
allows to conclude that values of $m_{H^\pm} = O(m_W)$ can be
excluded. Such values are, however, still allowed in other
2HDMs.

These issues are illustrated in Sec.~3.3, after defining in Sec.~3.1 the
class of 2HDMs considered, and presenting the NLO corrections at the
amplitude level in Sec.~3.2. 

%
\subsection{Couplings of Higgs bosons to fermions}

\noindent 
Models with $n$ Higgs doublets have generically a Yukawa 
Lagrangian (for the quarks) of the form:
\begin{equation} 
 - h^d_{ij} \,{\overline{q'}_L}_i \,\phi_1 \, {d'_R}_j 
 - h^u_{ij} \,{\overline{q'}_L}_i \,{\widetilde \phi}_2 \,{u'_R}_j 
 +{\rm h.c.}\,,
\label{yukpot}
\end{equation}
where $q'_L$, $\phi_i$, ($i=1,2$) are SU(2) 
doublets (${\widetilde \phi_i} = i \sigma^2 \phi_i^*$); 
$u'_R$, $d'_R$ are SU(2) singlets
and 
$h^d$, $h^u$ denote $3\times3$ Yukawa matrices. 
To avoid 
flavour changing neutral couplings at the 
tree--level, it is sufficient to 
impose that no more than one Higgs doublet couples to the same 
right--handed field, as in eq. (\ref{yukpot}). 

After a rotation of the quark fields from the current eigenstate to
the mass eigenstate basis, and an analogous rotation of the charged
Higgs fields through a unitary $n \times n$ matrix $U$, we assume 
that only one of the $n-1$ charged physical Higgs bosons
is light enough
to lead to observable effects in low energy processes. 
The $n$--Higgs doublet model then reduces to a generalized
2HDM, with 
the following Yukawa interaction for this 
charged physical Higgs boson denoted by $H^+$:
\begin{equation} 
\frac{g}{\sqrt{2}} \left\{
      X \,{\overline{u}_L} V \frac{m_d}{m_W} \, {d_R}+
      Y \,{\overline{u}_R} 
\frac{m_u}{m_W} V  \, d_L \right\} H^+ 
\label{higgslag}
\end{equation}
In~(\ref{higgslag}), $V$ is the Cabibbo--Kobayashi--Maskawa matrix
and the symbols $X$ and $Y$
are defined in terms of elements of 
the matrix $U$ (see citations in ref. \cite{BG98}). 
Notice that $X$ and $Y$ are
in general complex numbers and therefore potential sources of
CP violating effects. 
The ordinary Type~I and Type~II 2HDMs (with $n=2$), are
special cases of this generalized class, with 
$(X,Y) = (- \cot \beta, \cot \beta)$
and
$(X,Y) = ( \tan \beta, \cot \beta)$, respectively.

\subsection{NLO corrections at the amplitude level}
\noindent 
It turns out that the charged Higgs contributions 
do not induce operators in addition to those in the SM 
Hamiltonian $H_{eff}^W$ in eq. (\ref{heffW}). More specifically,
only step {\it 1)} below eq. (\ref{operators}),
gets modified when adding the
charged Higgs boson contributions to the SM one.
The new contributions to the matching conditions have been
worked out independently by several groups~\cite{CRS,Giudice97,BG98}, 
by simultaneously integrating out 
all heavy particles, $W$, $t$, and $H^+$ at the scale
$\mu_W$. This is a reasonable 
approximation provided $m_{H^\pm}$ is of the same order of magnitude
as $m_W$ or $m_t$.     

Indeed, the lower limit on $m_{H^\pm}$ coming from LEP~I, of 
$45\,$GeV, guarantees already $m_{H^\pm} = O(m_W)$. There 
exists a higher lower bound from LEP~II of 
$55\,$GeV for any value of $\tan \beta$~\cite{JANOT} 
for Type~I and Type~II models, which has been recently criticized in 
ref.~\cite{BD}. This criticism is based on the fact that 
there is no lower
bound on $m_A$ and/or $m_H$ coming from LEP~\cite{KRA,Maettig}.

After performing steps {\it 1)}, {\it 2)}, and {\it 3)}
listed below eq. (\ref{operators}), it is 
easy to obtain the quark level amplitude
$A(b \to s \gamma)$.
As  the matrix elements $\bra s \gamma|O_i|b \ket$ are proportional
to the tree--level matrix element of the operator $O_7$,
the amplitude $A$ can be written in the compact form
\beq
\label{ampl}
A(b \to s \gamma) = \frac{4 G_F}{\sqrt{2}} V_{tb} V_{ts}^* \,
\overline{D} \,
\bra s \gamma| O_7 |b\ket_{tree} \quad .
\eeq
For the following discussion it is useful to decompose the reduced amplitude 
$\overline{D}$
in such a way that the dependence on the couplings $X$ and $Y$
(see eq. (\ref{higgslag}))
becomes manifest:
\begin{equation}
\label{dsplit}
\overline{D} 
= \overline{D}_{\rm SM} + 
X Y^* \overline{D}_{\rm XY} + 
|Y|^2 \overline{D}_{\rm YY} \quad .
\end{equation}
\FIGURE{\epsfig{file=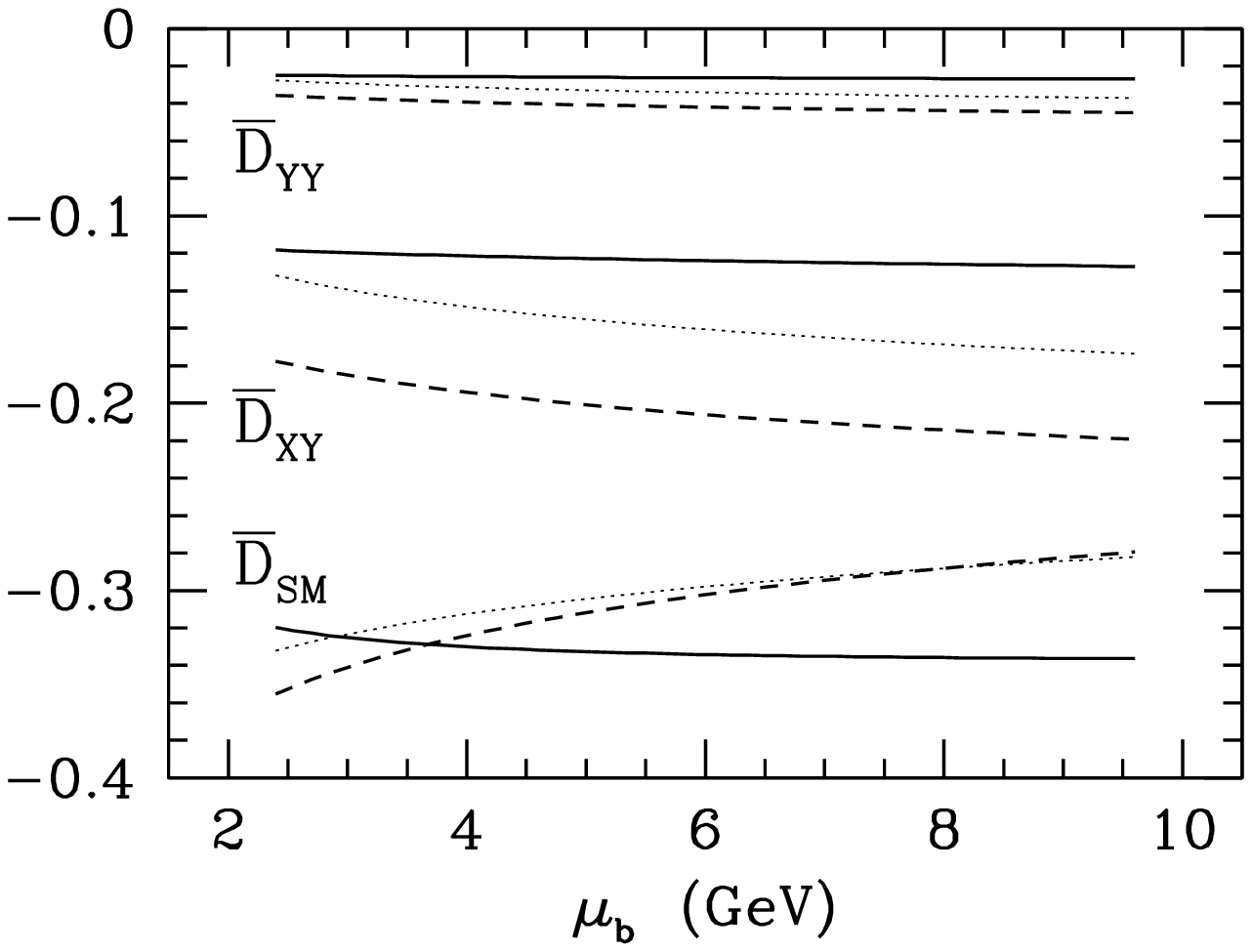,height=5.00cm,width=7.00cm}%
\caption{LO (dashed) and NLO (solid) predictions of the various pieces
of the reduced amplitude $\overline{D}$ for $m_{H^\pm} =100\,$GeV (see text).}
        \label{fig:radk}}
In Fig. \ref{fig:radk} 
the individual $\overline{D}$ quantities are shown in 
LO (dashed) and NLO (solid) 
order, for $m_{H^\pm} = 100$ GeV, as a function of $\mu_b$; 
all the other input parameters
are taken at their central values, as specified in ref. \cite{BG98}.  
To explain the situation, one can concentrate on the curves for 
$\overline{D}_{\rm XY}$.
Starting from the LO curve (dashed), the final NLO prediction is due
to the change of the Wilson coefficient $C_7$, shown by 
the dotted curve, and by the inclusion of the virtual
QCD corrections to the matrix elements. This results into a further
shift from the dotted curve to the solid curve. Both effects
contribute with the same sign and with similar magnitude, as it 
can be seen in Fig.~\ref{fig:radk}. 
The size of the NLO corrections in the term $\overline{D}_{\rm XY}$ in
(\ref{dsplit}) is
\begin{equation}
\frac{\Delta \overline{D}_{\rm XY}}{\overline{D}_{\rm XY}^{LO}} \equiv
\frac{\overline{D}_{\rm XY}^{NLO}-\overline{D}_{\rm
    XY}^{LO}}{\overline{D}_{\rm XY}^{LO}}
\sim - 40\% \,!
\end{equation}
A similarly large correction is also obtained for
$\overline{D}_{\rm YY}$. For the SM contribution $\overline{D}_{\rm SM}$,
the situation is different: the corrections to the
Wilson coefficient $C_7$ and the corrections due to the virtual
corrections in the matrix elements are smaller individually, 
and furthermore tend to cancel when combined, as shown in 
Fig. \ref{fig:radk}

The size of the corrections in $\overline{D}$ strongly depends on the 
couplings $X$ and $Y$ (see eq. (\ref{dsplit})):
$\Delta \overline{D}/\overline{D}$ is small, if the SM dominates, but it 
can reach values such as $-50\%$ or even worse, if the SM and 
the charged Higgs
contributions are similar in size but opposite in sign. 

\subsection{Results}
The branching ratio $BR(B \to X_s \gamma)$ can be sche- matically written as
\begin{equation}
\label{schematic}
BR(B \to X_s \gamma)  \propto |\overline{D}|^2 + \cdots 
\quad ,
\end{equation}
where the ellipses stand for Bremsstrahlung contributions, electroweak
corrections and nonperturbative effects.
As required by perturbation theory,
$|\overline{D}|^2$ in eq. (\ref{schematic}) should
be understood as
\begin{equation}
\label{dsq}
|\overline{D}|^2 = |\overline{D}^{LO}|^2 \left[ 1 + 2 \mbox{Re} \left( 
\frac{\Delta \overline{D}}{\overline{D}^{LO}} \right)
\right] \quad ,
\end{equation}
i.e., the term $|\Delta \overline{D}/\overline{D}^{LO}|^2$ is omitted.
If $\mbox{Re}(\Delta \overline{D}/$ $\overline{D}^{LO})$ 
is larger than $50\%$ in magnitude
and negative, the NLO branching ratio becomes  negative, i.e. the 
truncation of the perturbative series at the NLO level is not 
adequate for the corresponding couplings $X$ and $Y$.
As shown in ref.~\cite{BG98}, 
this can happen even for modest 
values of $X$ and $Y$.

However,
theoretical predictions for the branching ratio in 
Type~II models stand, in general, on a rather solid ground.
Fig.~\ref{fig:brIIcg} shows the low--scale 
dependence of $BR(B \to X_s \gamma)$ for matching scale $\mu_W = m_{H^\pm}$, for 
$m_{H^\pm}>100\,$GeV. It is less than $\pm 10\%$ for any value of 
$m_{H^\pm}$ above the LEP~I lower bound of $45\,$GeV. 
\FIGURE{\epsfig{file=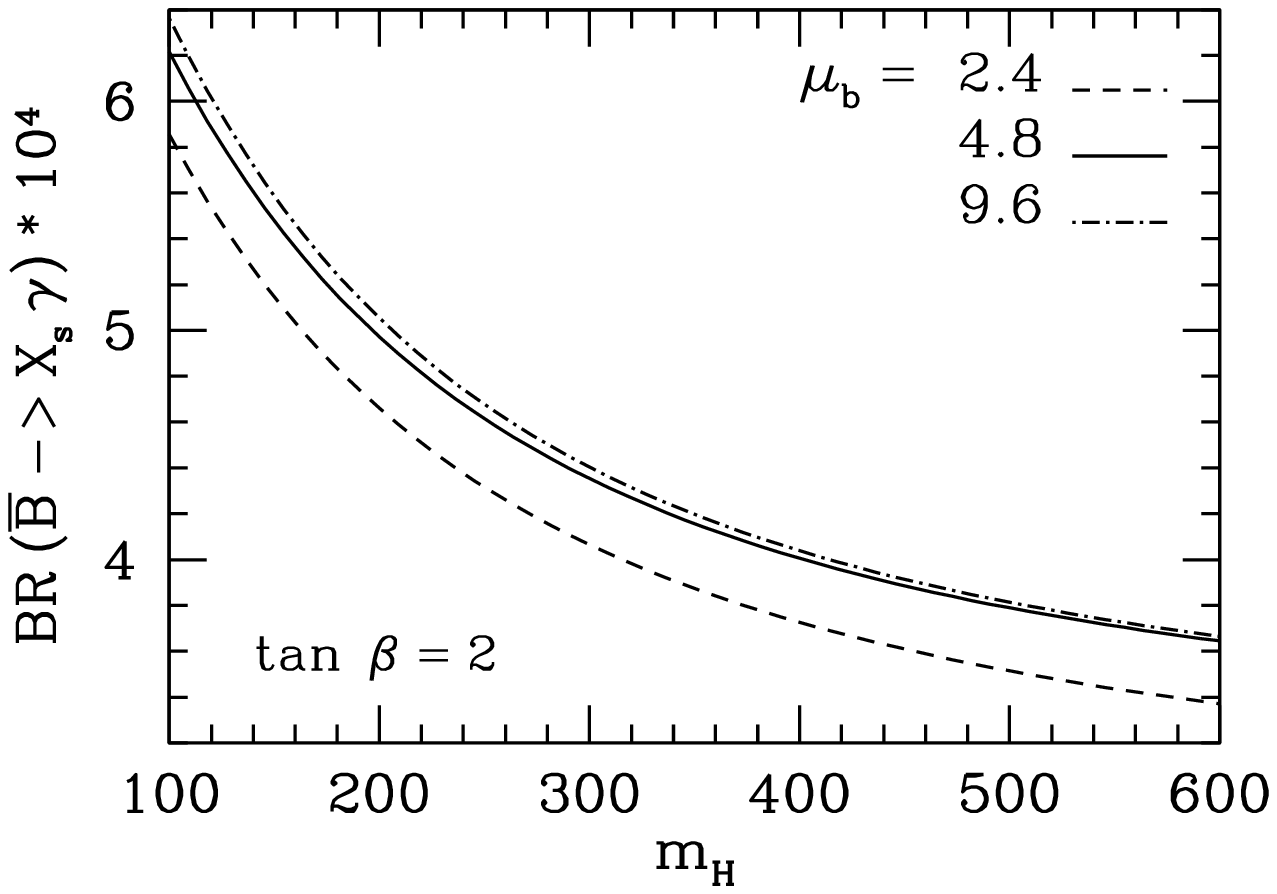,height=5.00cm,width=7.00cm}%
\caption{$BR(B \to X_s \gamma)$ in a Type~II model with $\tan \beta =2$, for
various values of $\mu_b$. The leading QED corrections are
included (see text).}
        \label{fig:brIIcg}}
Such a small scale 
uncertainty 
is a generic feature of Type~II models and remains true for 
values of $\tan \beta$ as small as $0.5$.
In this, as in the following figures where reliable NLO 
predictions 
are presented, the recently calculated leading QED 
corrections are included in the way
discussed in the addendum~\cite{BG98add} of
ref.~\cite{BG98}. They are not contained in the result
shown in Fig.~\ref{fig:radk}, 
which has  an illustrative aim only. 

In Type~II models, the theoretical estimate of $BR(B \to X_s \gamma)$ 
can be well above the experimental upper bound of 
$4.5 \times 10^{-4}$ ( $95\%$ C.L.) \cite{CLEOneu}, 
leading to constraints in the $(\tan \beta, m_{H^\pm})$
plane. The region excluded by the 
CLEO bound, as well as by other   
hypothetical experimental bounds, is given in Fig.~\ref{fig:contour}.
\FIGURE{\epsfig{file=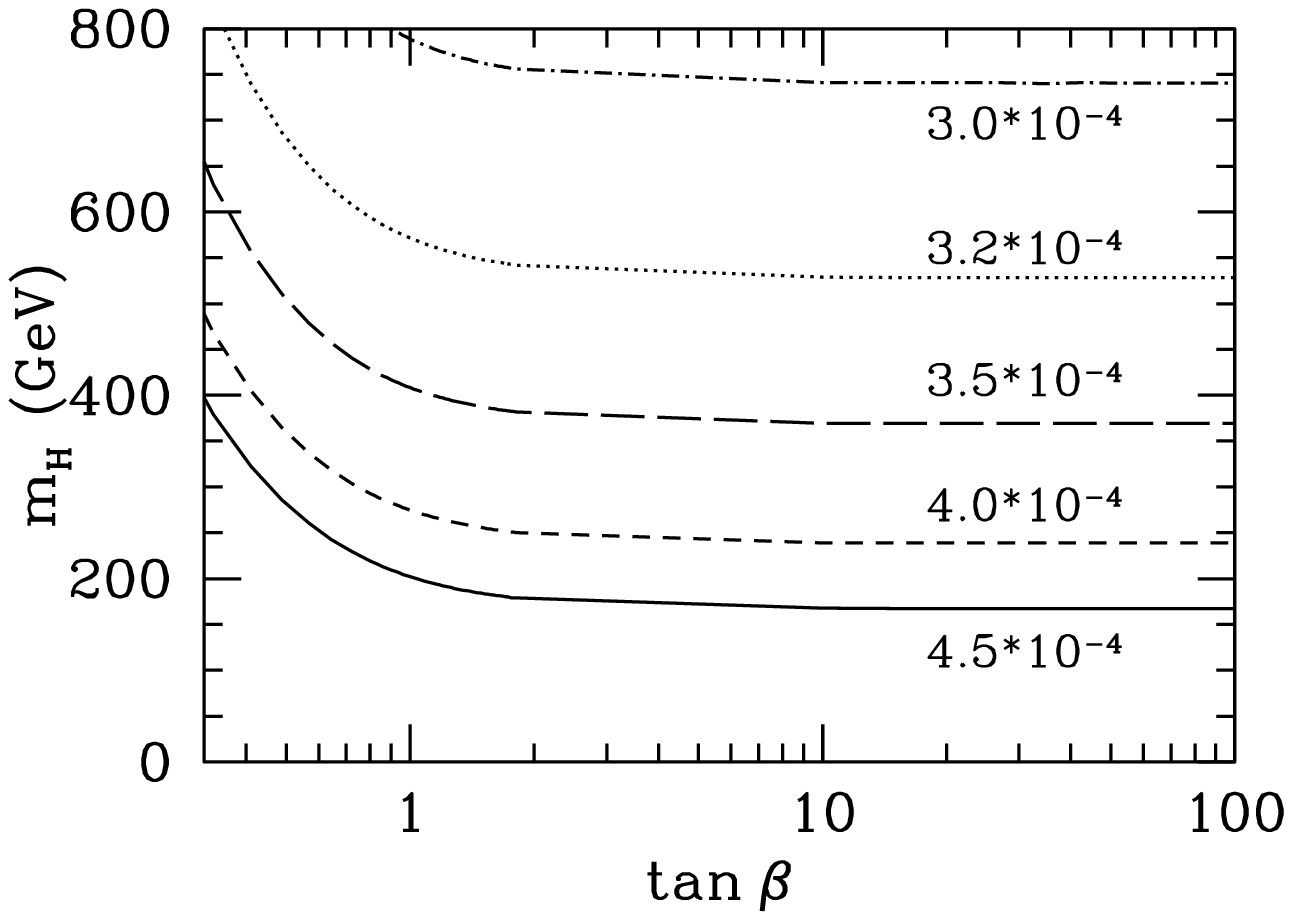,height=5.00cm,width=7.00cm}%
\caption{Contour plot in $(\tan \beta,m_{H^\pm})$ in Type~II models, 
obtained by using the
NLO expression for $BR(B \to X_s \gamma)$ 
and possible experimental upper bounds.
The leading QED corrections are included. The 
allowed region is above the corresponding curves.}
        \label{fig:contour}}
For $\tan \beta = 0.5$,1,5, we exclude respectively 
$m_{H^\pm} \le 280$, 200, 170 GeV, using the present upper bound from
CLEO.

Also in the case of complex couplings, the results for 
$BR(B \to X_s \gamma)$ range from ill--defined, to uncertain, up to 
reliable. One particularly interesting case in which 
the perturbative expansion can be safely truncated at the NLO level,
is identified by: 
$Y=1/2$, $X = 2 \exp (i\phi)$,  and $m_{H^\pm}=100\,$GeV. 
The corresponding branching ratios, shown in Fig.~\ref{fig:brok}, 
are consistent with the 
CLEO measurement, even for a relatively small value of $m_{H^\pm}$
in a large range of $\phi$. 
Such a light charged Higgs can contribute to 
the decays of the $t$--quark, through the mode $t \to H^+ b$.
\FIGURE{\epsfig{file=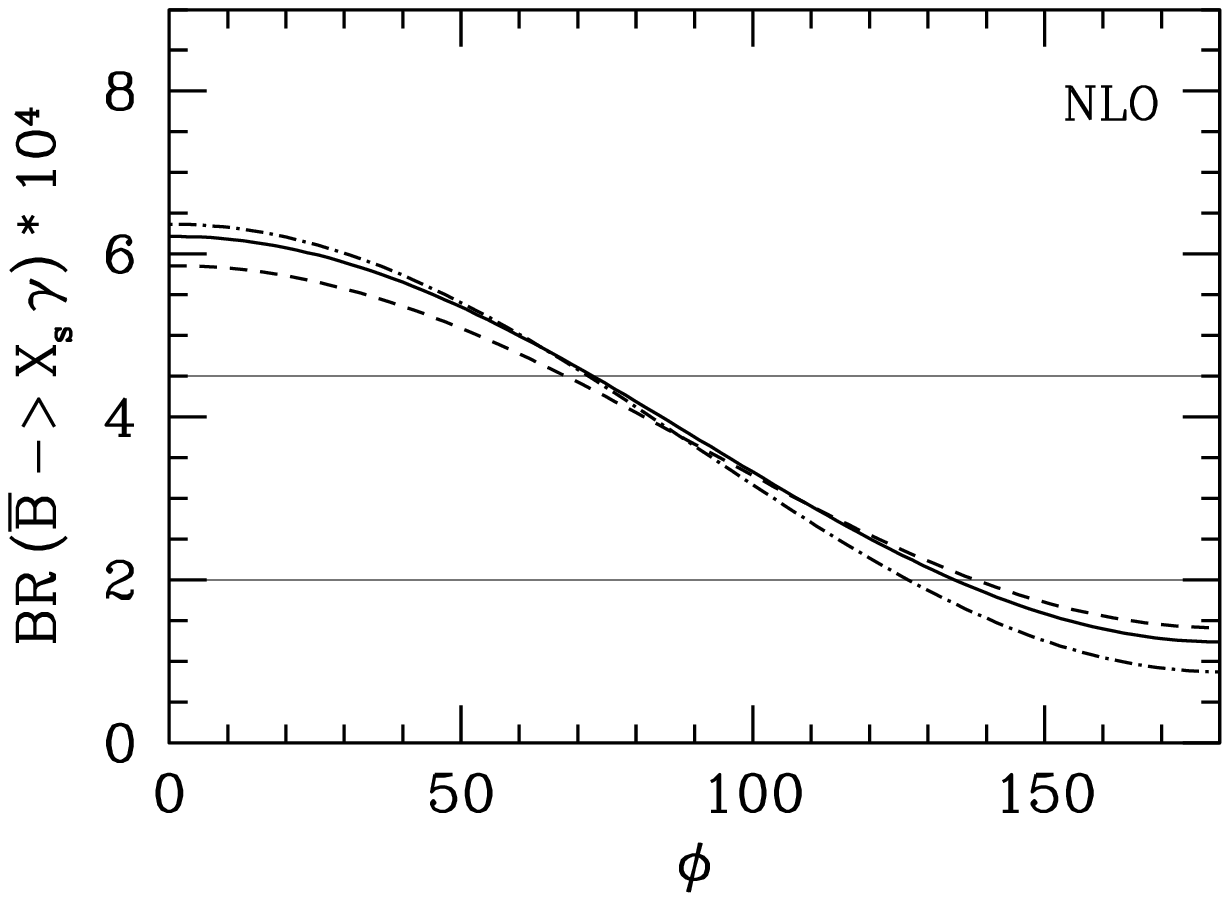,height=5.00cm,width=7.00cm}%
\caption{$BR(B \to X_s \gamma)$ 
as a function of $\phi$, where $\phi$ parametrizes
$X=2 \exp(i\phi)$, for $Y=1/2$, $m_{H^\pm}=100$ GeV. Solid, dashed
and dash--dotted lines correspond to $\mu_b=4.8,2.4,9.6$ GeV.
The leading QED corrections are included (see text).
Superimposed is the range of values allowed by the CLEO measurement.}
        \label{fig:brok}}

The imaginary parts in the $X$ and 
$Y$ couplings induce --together with the absorptive parts
of the NLO loop-functions-- CP rate asymmetries in $B \to X_s \gamma$.
A priori, these can be expected to be large. We find, however, 
that choices of the couplings $X$ and $Y$ which render the 
branching ratio stable, induce in general small asymmetries, not
much larger than the modest value of $1\%$ obtained in the SM \cite{AAG98}. 

\section{$B \to X_s \gamma$ in SUSY models}
Rare decays also  
provide guidelines for supersymmetry model building.   
Their  observation, or the upper limits set on them, yields  
stringent constraints on the many parameters 
in the soft supersymmetry-breaking terms. The processes
involving transitions between first and second generation quarks,
namely FCNC processes in the $K$ system, are considered to be most 
efficient in shaping viable supersymmetric flavour models. 

The severe experimental constraints on flavour violations have no direct 
explanation in the structure of the MSSM. This is the 
essence of the well-known supersymmetric flavour problem.
There exist several supersymmetric models (within the MSSM) with 
specific solutions to this problem. Most popular are the ones  
in which the dynamics of flavour sets in above the supersymmetry 
breaking scale and the flavour problem is killed by the 
mechanisms of communicating supersymmetry breaking to the 
experimentally accessible sector: In the constrained minimal supersymmetric 
standard model (mSUGRA), supergravity is the mediator between the 
supersymmetry breaking and the visible sector \cite{MSUGRA}. 
In  gauge mediated 
supersymmetry  breaking models (GMSBs), the communication 
between the two sectors is realized by gauge interactions \cite{GMSBs}.
More recently, the anomaly mediated 
supersymmetry breaking models (AMSBs) were proposed, in which the  
two sectors  are linked by interactions suppressed by the Planck mass
\cite{ANOMAL}. 
Furthermore, there are other classes of models in which the flavour problem is 
solved by particular flavour symmetries. 

Neutral flavour transitions involving the b quark,
do not pose yet serious threats to these
models. Nevertheless, the decay $B \to X_s \gamma$ has
already carved out some regions in the space of free
parameters of most of the models in the above classes (see \cite{THEO} 
and references therein). In particular, it dangerously constrains  
several somewhat tuned realizations of these
models \cite{TUNED}. Once the experimental precision is 
increased, this decay will undoubtedly help 
selecting the viable regions of the parameter space in the above 
class of models and/or discriminate among these or other possible
models. It is therefore important to calculate the rate of this 
decay as precisely as possible, for generic supersymmetric models. 

As we saw in section 2, $BR(B \to X_s \gamma)$ is known up to
NLO precision in the SM.
The calculation of this branching ratio within general supersymmetric models 
is still far from this level of sophistication.
There are several contributions to the decay amplitude: 
Besides the SM- and the charged Higgs one,
there are also
chargino-, gluino- and neutralino contributions.
All these were calculated in \cite{FRA}
within the mSUGRA model.  
The inclusion of LO QCD corrections 
was assumed to follow the SM pattern. 
A calculation taking into account solely
 the gluino contribution has been performed 
in \cite{MAS} for a generic supersymmetric model, but no QCD
corrections were included. 

An interesting NLO analysis of $B \to X_s \gamma$ was recently performed 
\cite{Giudice98} in a specific class of 
models where the only source of flavour violation 
at the electroweak scale is 
encoded in the CKM matrix.
The calculations were done in the limit
\begin{eqnarray}
\mu_{\tilde{g}} &\equiv& {\cal O}(m_{\tilde{g}},
m_{\tilde{q}},
m_{\tilde{t}_1}) \gg \mu_W \nonumber \\
                &\equiv& {\cal O}(m_W,m_{H^\pm},m_{\chi^\pm},m_{\tilde{t}_2}),
\end{eqnarray}
and
terms of order $(\mu_W/\mu_{\tilde{g}})^p$ ($p\ge 2$) were discarded.
At the scale $\mu_W$ the new contributions can be matched onto the same
operators as in the SM (see also ref. \cite{Misiak99}). 
It is shown in the analysis \cite{Giudice98} that much lower values for
$m_{H^\pm}$ are allowed
than in the type-II 2HDM, discussed in section 2, due to the possiblity of
destructive
interference between the charged Higgs and the chargino contributions.
It is illustrated in Fig.~\ref{fig:giudice}  that in such a 
cancellation scenario the NLO QCD corrections
are important: The uppermost curve is the LO result
for the type-II 2HDM with $m_{H^\pm}=100$ GeV; switching on 
also the chargino contribution at LO (using the parameters mentioned in the
caption), leads to the second curve (from the bottom). 
\FIGURE{\epsfig{file=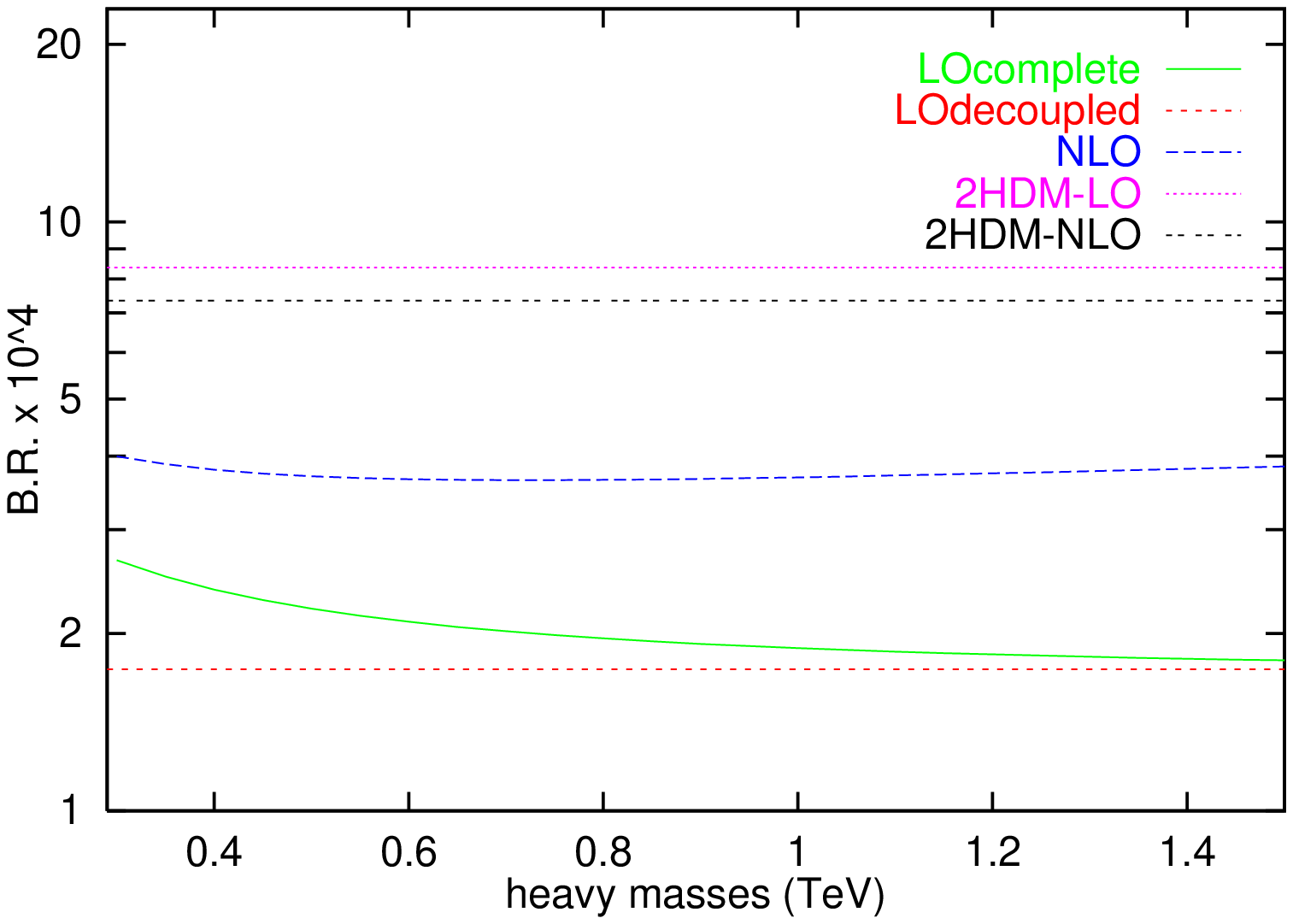,height=5.00cm,width=7.00cm}%
\caption{$BR(B \to X_s \gamma)$ as a function of $\mu_{\tilde{g}}$
for the parameters: $\tan \beta =1$, $m_{H^\pm} = m_{\tilde{t}_2} =
m_{\chi_2} = 100$ GeV, $m_{\chi_1} = 300$ GeV, 
$\theta_{\tilde{t}}=-\pi/10$, $A_b=A_t$; all
heavy particle masses equal to $\mu_{\tilde{g}}$; the lighter chargino is 
predominantly higgsino. Figure taken from M. Ciuchini et al. \cite{Giudice98}
        \label{fig:giudice}}}
The NLO result
for the combination of the charged-Higgs and the chargino contribution 
is represented by the middle curve.   

This calculation, however,
cannot be used in particular directions of the parameter 
space of the above listed models in which quantum effects induce a 
gluino contribution as large as the chargino or the SM contributions. 
Nor it can be used as a model-discriminator tool, able to constrain 
the potentially
large sources of flavour violation typical of generic 
supersymmetric models.  

The flavour non diagonal vertex gluino-quark-squark induced by
the flavour violating scalar mass term and trilinear terms
is particularly interesting. This is generically assumed to induce the 
dominant contribution to quark flavour transitions, as this vertex is weighted
by the strong coupling constant $g_s$.
Therefore, it is often taken as the 
only contribution 
to these transitions and in particular to the $B \to X_s \gamma$
decay, when extracting order of magnitude upper bounds
on flavour violating terms in the scalar potential \cite{MAS,HAG}.
Once the constraints coming from the experimental measurements are imposed, 
however, the gluino contribution is reduced to values such that the SM 
and the other supersymmetric contributions cannot be neglected 
anymore. Any LO and NLO calculation of the $B \to X_s \gamma$
rate in generic supersymmetric models, therefore, should then include
all possible contributions. 

The gluino contribution, however, presents some peculiar features related
to the implementation of the QCD corrections.
In ref. \cite{BGHW99} this contribution to the decay $b \rightarrow s \gamma$
is therefore investigated in great detail for
 supersymmetric models with generic soft terms.
It is shown
that 
the relavant operator basis of the SM effective Hamiltonian gets enlarged 
to contain magnetic and chromomagnetic operators with an extra factor of 
$\alpha_s$
and weighted by a quark mass $m_b$ or $m_c$, and also 
magnetic and chromomagnetic operators of lower dimensionality, as well as 
additional scalar and tensorial 
four-quark operators.
A  few results of the analysis in ref. \cite{BGHW99} are 
given  in the following, showing the effect of the 
LO QCD corrections on constraints on supersymmetric sources of 
flavour violation.

To understand the sources of flavour violation which may be present in
supersymmetric models in addition to those enclosed in the CKM matrix,
one has to consider the contributions to the squark mass matrices
\begin{equation}
{\cal M}_{f}^2 =  
\left( \begin{array}{cc}
  m^2_{f,LL}   & m^2_{f,LR} \\
  m^2_{f,RL}  &  m^2_{f,RR}                 
 \end{array} \right) +
\label{squarku}
\nonumber
\end{equation}
\begin{equation}
  \left( \begin{array}{cc}
  F_{f,LL} +D_{f,LL} &  F_{f,LR} \\
 F_{f,RL} & F_{f,RR} +D_{f,RR}                
 \end{array} \right) \quad ,
\nonumber
\label{squarku2}
\end{equation}
where $f$ stands for up- or down-type squarks.
In the super  CKM basis where the quark mass matrices are diagonal 
and the squarks are rotated in parallel to their superpartners,
the $F$ terms  from the superpotential and the $D$ terms 
turn out to be diagonal 
$3 \times 3$ submatrices of the 
$6 \times 6$
mass matrices ${\cal M}^2_f$. This is in general not true 
for the additional terms (\ref{squarku}), originating from  the soft 
supersymmetry breaking potential. As a consequence, 
gluino contributions to the
decay $b \to s \gamma$ are induced by the off-diagonal
elements of the soft terms 
$m^2_{f,LL}$, $m^2_{f,RR}$, $m^2_{f,LR}$ and $m^2_{f,RL}$.

It is convenient to select one possible 
source of flavour violation in the squark sector at a time and
assume that all the remaining ones are vanishing. Following
ref.~\cite{MAS}, all diagonal entries in 
$m^2_{\,d,\,LL}$, $m^2_{\,d,\,RR}$, and $m^2_{\,u,\,RR}$
are set to be equal and their common value is denoted by
$m_{\tilde{q}}^2$.  The branching ratio can then be studied as a
function of 
\begin{equation} 
\delta_{LL,ij} = \frac{(m^2_{\,d,\,LL})_{ij}}{m^2_{\tilde{q}}}\,, 
\hspace{0.1truecm}
\delta_{RR,ij} = \frac{(m^2_{\,d,\,RR})_{ij}}{m^2_{\tilde{q}}}\,, 
\hspace{0.1truecm} 
\label{deltadefa}
\end{equation}
\begin{equation} 
\delta_{LR,ij} = \frac{(m^2_{\,d,\,LR})_{ij}}{m^2_{\tilde{q}}}\,, 
\, (i \ne j).
\label{deltadefb}
\end{equation}
The remaining crucial parameter needed to determine the 
branching ratio is $x = m^2_{\tilde{g}}/m^2_{\tilde{q}}$,
where $m_{\tilde{g}}$ is the gluino mass.
In the following, we concentrate on the LO QCD corrections to the 
gluino contribution. 
\FIGURE{\epsfig{file=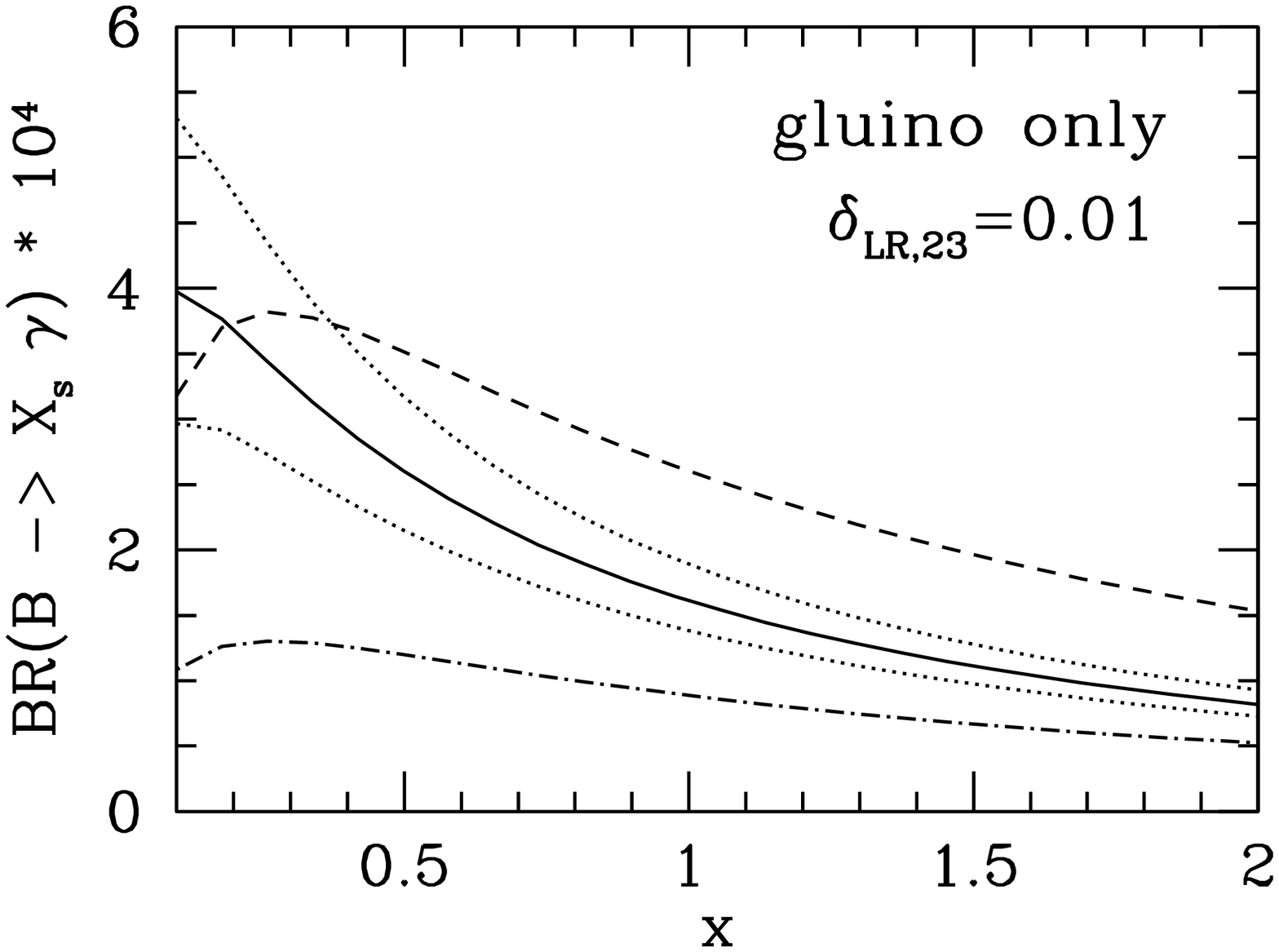,height=5.00cm,width=7.00cm}%
\caption{Gluino-induced branching ratio $BR(B \to X_s \gamma)$ 
 as a function of $x= m^2_{\tilde{g}}/m^2_{\tilde{q}}$, obtained when
 the only source of flavour violation is $\delta_{LR,23}$ (see text).}
        \label{fig:sizeqcd23lr}}
\FIGURE{\epsfig{file=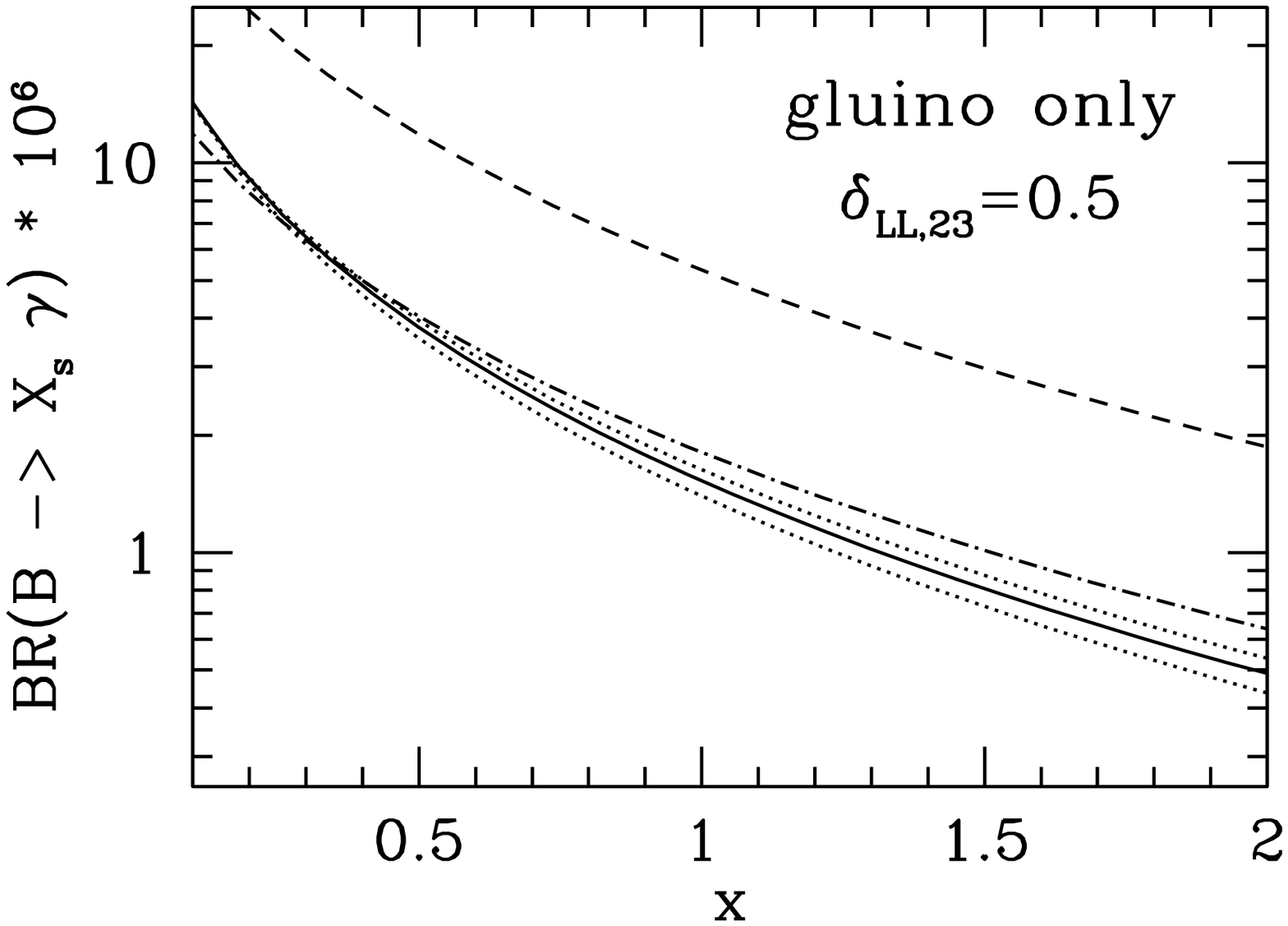,height=5.00cm,width=7.00cm}%
\caption{Same as in Fig.~\ref{fig:sizeqcd23lr} when only
 $\delta_{LL,23}$ is non-vanishing.}
        \label{fig:sizeqcd23ll}}
In Figs.~\ref{fig:sizeqcd23lr} and \ref{fig:sizeqcd23ll},
the solid lines show  the QCD corrected
branching ratio, when only $\delta_{LR,23}$ or
$\delta_{LL,23}$ are non vanishing.  
The branching ratio is plotted as a function of
$x$, using
$m_{\tilde{q}}=500\,$GeV.  The  dotted lines show the range of variation
of the branching ratio, when the
renormalization scale $\mu_b$ varies in the interval $2.4$--$9.6\,$GeV. 
Numerically, the scale uncertaintly in
$BR(B \to X_s \gamma)$ is about
$\pm 25\%$. An extraction of bounds on the $\delta$
quantities more precise than just an order of magnitude, therefore,
would require the inclusion of NLO QCD
corrections. It should be noticed, however, that the inclusion of the
LO QCD corrections has already removed the large ambiguity on the
value to be assigned to the factor $\alpha_s(\mu)$ in the
gluino-induced operators. Before
adding QCD corrections, the scale in this factor can assume all values
from $O(m_b)$ to $O(m_W)$: the difference between $BR(B \to X_s \gamma)$
obtained when $\alpha_s(m_b)$ or when $\alpha_s(m_W)$ is used, is
of the same order as the LO QCD corrections.  The corresponding values
for $BR(B \to X_s \gamma)$ for the two extreme choices of $\mu$ are
indicated in Figs.~\ref{fig:sizeqcd23lr} and~\ref{fig:sizeqcd23ll} by
the dot-dashed lines ($\mu=m_W$) and the dashed lines
($\mu=m_b$). The
choice $\mu = m_W$ gives values for the non-QCD corrected 
$BR(B \to X_s \gamma)$ relatively close to the band obtained when the
LO QCD corrections are included, if only $\delta_{LL,23}$ is
non-vanishing. Finding a corresponding value of $\mu$ that minimizes
the QCD corrections in the case studied in Fig.~\ref{fig:sizeqcd23lr},
when only $\delta_{LR,23}$ is different from zero,
depends strongly on the value of $x$.
In the context of the full LO result, it is important
to stress that the explicit $\alpha_s$ factor 
                       has to be evaluated 
- like the Wilson coefficients - at a scale $\mu=O(m_b)$.

In spite of the large uncertainties which the branching ratio 
$BR(B \to X_s \gamma)$ still has at  LO in QCD, it is possible
to extract indications on the size that the $\delta$-quantities 
may maximally acquire without inducing conflicts with the 
experimental measurements (see \cite{BGHW99}). 

\section{Summary}
Significant progress in the theoretical description of rare B decays has been 
achieved during the last few years. NLO QCD corrections are available
for radiative inclusive decays in the SM. Power
correction ($1/m_b^2$, $1/m_c^2$) are also under control. The description
of the Fermi motion of the $b$ quark in the $B$ meson has been refined. 
Important NNLO QCD improvement
of the matching conditions of the operators relevant for $B \to X_s \ell^+
\ell^-$ was made, which removes the $\pm 16\%$ matching scale uncertainty in
the invariant mass distribution of the lepton pair.
In some regions of the parameter space,
NLO QCD corrections to $BR(B \to X_s \gamma)$ in 2HDMS are huge. They are,
however, under control in the type-II model and are important to derive
reliable bounds on $\tan \beta$ and $m_{H^\pm}$. NLO QCD corrections
to $BR(B \to X_s \gamma)$ are available also in particular SUSY scenarios.
Calculations for more general situations are in progress.

\acknowledgments{I thank A. Ali, H. Asatrian, F. Borzumati, T. Hurth,
T. Mannel,
and D. Wyler for pleasant collaboration on many topics presented in this 
talk. I also thank P. Liniger for his help to implement the figures.} 


\end{document}